\documentclass[aps,prl,preprint,showpacs,superscriptaddress]{revtex4}
\usepackage{bm,amsmath,amssymb,latexsym,mathrsfs,enumerate}
\usepackage[mathcal]{euscript}
\usepackage{graphicx, subfigure}

\begin{document}

\title{On the Temperature Dependence of the Casimir Force for Bulk Lossy Media}

\author{V.~A.~Yampol'skii}
\affiliation{Advanced Science Institute, The Institute of Physical
and Chemical Research (RIKEN), Wako-shi, Saitama, 351-0198, Japan}
\affiliation{A. Ya. Usikov Institute for Radiophysics and
Electronics National Academy of Sciences of Ukraine, 61085
Kharkov, Ukraine}
\author{Sergey Savel'ev}
\affiliation{Advanced Science Institute, The Institute of Physical
and Chemical Research (RIKEN), Wako-shi, Saitama, 351-0198, Japan}
\affiliation{Department of Physics, Loughborough University,
Loughborough LE11 3TU, UK}
\author{Z. A. Mayselis}
\affiliation{Advanced Science Institute, The Institute of Physical
and Chemical Research (RIKEN), Wako-shi, Saitama, 351-0198, Japan}
\affiliation{A. Ya. Usikov Institute for Radiophysics and
Electronics National Academy of Sciences of Ukraine, 61085
Kharkov, Ukraine}
\author{S. S. Apostolov}
\affiliation{Advanced Science Institute, The Institute of Physical
and Chemical Research (RIKEN), Wako-shi, Saitama, 351-0198, Japan}
\affiliation{A. Ya. Usikov Institute for Radiophysics and
Electronics National Academy of Sciences of Ukraine, 61085
Kharkov, Ukraine}
\author{Franco Nori}
\affiliation{Advanced Science Institute, The Institute of Physical
and Chemical Research (RIKEN), Wako-shi, Saitama, 351-0198, Japan}
\affiliation{Department of Physics, Center for Theoretical
Physics, Applied Physics Program, Center for the Study of Complex
Systems, The University of Michigan, Ann Arbor, MI 48109-1040, USA}

\begin{abstract}
We discuss the limitations of the applicability of the Lifshitz formula to describe the temperature dependence of the Casimir force between two bulk lossy metals. These limitations follow from the finite sizes of the interacting bodies. Namely,  Lifshitz's theory is not applicable when the characteristic wavelengths of the fluctuating fields, responsible for the temperature-dependent terms in the Casimir force, is longer than the sizes of the samples. As a result of this, the widely discussed linearly decreasing temperature dependence of the Casimir force can be observed only for dirty and/or large metal samples at high enough temperatures. This solves the problem of the inconsistency between the Nernst theorem and the ``linearly decreasing temperature dependence'' of
the Casimir free energy, because  this
linear dependence is not valid when $T \rightarrow 0$.

\end{abstract}

\date{\today}



\maketitle

\section{Introduction}

The Casimir effect is one of the most interesting macroscopic manifestation of the zero-point vacuum oscillations of the quantum
electromagnetic field. This effect manifests itself as an
attractive force arising between two uncharged bodies due to the difference of the zero-point oscillation spectrum in the absence and in the presence of these bodies (see, e.g., Refs.~\cite{m1,m3,r1,r3}).

The Casimir effect has attracted considerable attention because of its
numerous applications in quantum field theory, atomic physics,
condensed matter physics, gravitation and
cosmology~\cite{m1,m3,r1,r3,f}. The noticeable progress in the
measurements of the Casimir force~\cite{exp} has opened the way
for various potential applications in nanoscience~\cite{nan},
particularly in the development of nano-mechanical
systems~\cite{m3,r3,nan}.

\subsection{Problems linked to Lifshitz's theory for the Casimir force}

In spite of intensive studies on the Casimir effect, it is
surprising that such an important problem as the temperature
dependence of this effect is still unclear and is still an issue
of lively discussion (see, e.g., Refs.~\cite{a-dr,dr,dr1,Hoy,prl,boris,svet}). The zero-temperature contribution to the force, originating from quantum fluctuations of the electromagnetic
field, is well understood. However, the contribution $F_{\rm
rad}(T)$ to the Casimir force originating from thermal fluctuations is a source of numerous controversies.

First, within the Lifshitz theory~\cite{2}, there is no continuous transition for the forces between ideal metals and real metals~\cite{dr}. The Lifshitz formula predicts an \emph{increase} of $F_{\rm
rad}(T)$ when increasing $T$ only for \emph{ideal} metals without relaxation. At the same time, for \emph{lossy} media with relaxation frequency $\nu \neq 0$, this formula gives a \emph{decrease} of $F_{\rm rad}(T)$ in a wide region of
temperatures. This decreasing term is related to the transparency of real metals for $s$-polarized (transverse electric) low-frequency fields. In other words, the behavior of $F_{\rm rad}(T)$
changes abruptly, in a jump-like manner, for \emph{infinitesimal}
$\nu$, in comparison to the case when $\nu = 0$. This discontinuous jump is not physical.

Second, the Casimir-Lifshitz entropy does not go to zero when $T \rightarrow 0$. This is unphysical, because it violates the Nernst theorem. This problem is still the focus of discussions (see, e.g., Ref.~\cite{Hoy}).

These obvious contradictions to common sense show that
some important physics is missing. Recently, Ref.~\cite{svet} indicated that the problems mentioned above can be solved if one takes into account the spatial dispersion of the low-frequency metal conductivity.

\subsection{Summary}

In this work, we demonstrate that there exist simple limitations for the applicability of Lifshitz's theory. We show that the Casimir force for good metals, where the plasma frequency $\omega_p$ is much higher than other characteristic frequencies of the system, \emph{increases monotonically} with temperature $T$ if the following inequality is satisfied:
\begin{equation}\label{1}
\nu \ll 2\pi  c/L.
\end{equation}
Here $L$ is the width of a sample. This condition (1) is typically satisfied for pure metals. For instance, it is valid for metal discs with area $\sim
10^{-2}$~cm$^2$, if the relaxation frequency is less than
$10^{12}$~s$^{-1}$.

\section{Analysis of the temperature dependence of the Lifshitz formula}

The purpose of this section is to show that the main contribution to the ``linearly decreasing with temperature term'' in the Casimir force between \emph{infinite} plates of lossy metals comes from the fluctuating fields with small frequencies
\begin{equation}\label{ineq}
 \omega \lesssim \nu.
\end{equation}
We analyze the Lifshitz expression for the Casimir force taken from Ref.~\cite{2} in the form of an integral over \emph{real} frequencies $\omega$. We use the Drude model for the permittivity $\varepsilon$,
\begin{equation}\label{eps}
\varepsilon(\omega)=1-\frac{\omega_p^2}{\omega(\omega+i\nu)}.
\end{equation}
In this case, the thermal term $F_{\mathrm{rad}}$ in the Casimir force per unit area can be written in the following form:
\begin{gather}
F_{\mathrm{rad}}=\frac{\hbar}{\pi^2
c^3}\Re\int_{0}^{\infty}\mathrm{d} \omega\int\mathrm{d}p\,
p^2\omega^3 \frac{1}{\exp(2\hbar\omega/kT)-1}
\notag\\\label{lif} \times\Bigg\{\left[\Big(\frac{s+p}{s-p}\Big)^2
\exp(-2i p\omega l/c)-1 \right]^{-1} +
\left[\Big(\frac{s+\varepsilon p}{s-\varepsilon p}\Big)^2 \exp(-2i
p\omega l/c)-1 \right]^{-1}\Bigg\}
\end{gather}
where
\begin{equation}\label{s}
s=\sqrt{\varepsilon(\omega)-1+p^2},
\end{equation}
$l$ is the separation between the interacting
bodies, and the symbol $\Re$ denotes the real part.
The integration trajectory over $p$ consists of two parts: from
$1$ to $0$ over the real axis, and from $i0$ to $+i\infty$ over the
imaginary axis.

We examine the difference $\Delta F_{\mathrm{rad}}$ between the contributions to
the Casimir force from thermal fluctuations for a dissipationless metal ($\nu=0$)
and for a metal with weak dissipation ($\nu\to0$),
\begin{equation}\label{s}
\Delta F_{\mathrm{rad}}= F_{\mathrm{rad}}\Big|_{\nu\to0}-
F_{\mathrm{rad}}\Big|_{\nu=0}.
\end{equation}
Namely,  $\Delta F_{\mathrm{rad}}$ describes the ``linearly decreasing with $T$'' part of the Casimir force $F_{\rm rad}(T)$ that appears in a jump-like manner at $\nu \neq 0$.
It is important to note that only the first term in the curly brackets in
Eq.~(\ref{lif}) [integrated over $p$ from $i0$ to $+i\infty$, and
over $\omega$ from $0$ to $+\infty$] produces this discontinuity. So, the difference $\Delta F_{\mathrm{rad}}$ can be written as
\begin{gather}
\Delta F_{\mathrm{rad}}=\frac{\hbar}{\pi^2
c^3}\Re\int_{0}^{\infty}\mathrm{d}\omega
\int_{0i}^{+i\infty}\mathrm{d}p\,
p^2\omega^3\frac{1}{\exp(2\hbar\omega/kT)-1}\notag\\\label{d_lif} \times \Bigg\{\left[\Big(\frac{s+p}{s-p}\Big)^2
\exp(-2i p\omega l/c)-1 \right]^{-1} -
\left[\Big(\frac{s|_{\nu=0}+p}{s|_{\nu=0}-p}\Big)^2 \exp(-2i
p\omega l/c)-1 \right]^{-1}\Bigg\} .
\end{gather}

Introducing the notation,
\begin{equation}\label{new_int}
t=\frac{\omega}{\nu}, \quad x=-\, \frac{2ip\omega l}{c}, \quad
\alpha=\frac{c}{2l\omega_p},
\end{equation}
and assuming that
\begin{equation}\label{ner}
\hbar\nu \ll kT,
\end{equation}
we obtain
\begin{gather}\label{fin}
\Delta F_{\mathrm{rad}}=-\frac{ kT}{8\pi^2
l^3}\Im\int_{0}^{\infty}\frac{\mathrm{d}t}{t}
\int_{0}^{+\infty}\frac{\mathrm{d}x\, x^2}{\bigg(\alpha
x+\sqrt{\alpha^2
x^2+\dfrac{t}{t+i}}\bigg)^4\Big(\dfrac{t+i}{t}\Big)^2
\mathrm{e}^x-1}
\end{gather}
where the symbol $\Im$ denotes the imaginary part.

It is seen from this equation that the difference $\Delta
F_{\mathrm{rad}}$ does \emph{not} depend on $\nu$, and that the main
contribution to this integral comes from $x\sim 1$ and $t\lesssim
1$. Thus, according to Eq.~(\ref{new_int}), \emph{the characteristic values of} $\omega$
\emph{are either of the order or less than} $\nu$.

For good metals with $\alpha\lesssim 1$, the value of the integral in Eq.~(\ref{fin}) is of the order of $1$, and
\begin{equation}\label{small}
 \Delta F_{\mathrm{rad}}\sim \frac{kT}{l^3} \qquad \left({\mathrm {for \,\,good \,\,metals \,\, with}} \,\,\alpha \lesssim 1\right).
\end{equation}
In the opposite case $\alpha\gg 1$, the characteristic frequencies are even smaller than $\nu$, $\omega \sim \nu/\alpha \ll \nu$. For such small distances $l$, the value of
the integral in Eq.~(\ref{fin}) is of the order of $1/\alpha^3$, and
\begin{equation}\label{large}
 \Delta F_{\mathrm{rad}}\sim \frac{kT \omega^3_p}{c^3} \qquad \left({\mathrm {when}}\,\, \alpha \lesssim 1; \,\,{\mathrm {i.e., \,\, when}}\,\,  l \ll \frac{c}{\omega_p}\right).
\end{equation}

\section{Limitations for the applicability of the Lifshitz formula and discussions}
Thus, the main contribution to the ``linearly decreasing with $T$ term'' $F_{\rm
rad}(T)$ in the Lifshitz theory comes from small frequencies
satisfying two inequalities,
\begin{equation}\label{2}
\omega \ll kT/\hbar, \qquad \omega \lesssim \nu.
\end{equation}
Obviously, the wavelengths of the fluctuating fields with such
frequencies should be much smaller than the size of the sample.
Otherwise, the sample \emph{cannot be considered as
semi-infinite}. In other words, the Lifshitz theory gives a
(\emph{physically} correct) $F_{\rm rad}(T)$ decreasing with
temperature  \emph{if}
\begin{equation}\label{3}
\nu \gg 2\pi  c/L
\end{equation}
and
\begin{equation}\label{4}
kT \gg 2\pi \hbar c/L.
\end{equation}
Moreover, it is clear that a metal with plasma frequency much
higher than other characteristic frequencies $\omega_i$, and with
$\nu \ll \omega_i$ (the frequency $2\pi c/L$ is among them),
should possess properties \emph{close to the ones of an ideal
metal}. This means, that the inequality (\ref{1}) ensures the
increase of the Casimir force with temperature, similarly to the
case of ideal metals.

Note that the condition (\ref{4}) did not hold in the experiment
in Ref.~\cite{3} (with a tiny metal sphere of radius
$R=151.3$~$\mu$m). For all temperatures used in that experiment,
the wavelengths of the fluctuating fields responsible for the
temperature decrease of the Casimir force (expected within the
Lifshitz theory with the Drude model for the permittivity) were of the order of $R$ and longer than the radius
$r \sim (Rl)^{1/2}$ of the effective interacting region of the
tiny sphere. Therefore, it is not surprising that the temperature
decrease of the Casimir force was not observed in Ref.~\cite{3}.

Note also that the condition in Eq.~(\ref{4}) solves the problem of the
inconsistency between the Nernst theorem and the linearly decreasing temperature dependence of the Casimir free energy. Indeed, the
linear asymptotic dependence of $F_{\rm rad}$ on $T$ is not
applicable when $T \rightarrow 0$.

\section{Acknowledgements}

We acknowledge partial support from the NSA, LPS, ARO, NSF grant No. EIA-0130383, JSPS-RFBR 06-02-91200, the EPSRC via No. EP/D072581/1, EP/F005482/1, and the ESF AQDJJ network programme.

\end{document}